\documentclass{article}
\usepackage{amsfonts}
\usepackage{amssymb}
\usepackage{amsmath}
\usepackage{amsthm}

\title{Canonical quantization \\ of motion on submanifolds}
\author{ Alexey V. Golovnev \\ {\small\it Arnold Sommerfeld Center for Theoretical Physics} \\ 
{\small\it Ludwig-Maximilians University}\\ 
{\small Theresienstr. 37, Munich, 80333 Germany} \\ 
{\small  Alexey.Golovnev@physik.uni-muenchen.de } }
\date{ }

\begin{document}

\maketitle
\begin{abstract}
This is an extended version of the talk given by the Author at 
the 40th Symposium on Mathematical Physics held in
Torun, Poland, June 25-28, 2008.
We review the methods of canonical quantization of free particle motion on
curved submanifolds considered as a system with second class constraints. The work is based on our
previous articles, \cite{myDirac} and \cite{mythin}. However, some new results are also 
presented.
\end{abstract}

\section{Introduction}
We consider the problem of quantum motion in curved spaces. It is well-known that in the case of Euclidean
spaces the correct quantum Hamiltonian is ${\hat H}=-\frac{{\hbar}^2}{2}\Delta$. Podolsky \cite{Podolsky} in 1928 proposed that
for arbitrary spaces it should be replaced by ${\hat H}=-\frac{{\hbar}^2}{2}{\Delta}_{LB}$ with ${\Delta}_{LB}$
being the Laplace-Beltrami operator. This postulate is a direct and geometrically clear generalization
of the dynamics in Euclidean spaces.
But if one wants to get the theory by some canonical procedure, he encounters a severe problem. For
any given classical theory there is an infinite number of quantum theories with a proper $\hbar\to 0$ limit.
Quantization is not unique. In Euclidean spaces Dirac recipe in Cartesian coordinates yields
experimentally correct result for the theories which we usually have in theoretical physics. However, in curved spaces we do not have a notion of Cartesian coordinates and can't
make a choice of the theory in this way. Of course, if the theory posesses a large enough symmetry it can be sometimes completely defined by the symmetry requirements directly at the quantum level, see \cite{SMP}. Nevertheless, we think it is very instructive to study the
properties of classical quantization methods for constrained systems since theories with constraints are
so important in modern physics. In particular, a possible solution for our problem is to embed the space under consideration into some Euclidean space
and to quantize the new theory as a theory with second class constraints. As we shall see below, the results depend both on the choice of embedding and on the method of quantization.

In Section 2 we describe the Dirac approach in the case of codimension 1 surfaces first, give an explicit operator
realization of it
and show that the Dirac procedure is ambiguous. The quantum Hamiltonian depends even on a particular form in which the equation
of surface is presented; a natural geometric way of fixing this freedom is explained. We also compare these results \cite{myDirac} with
those obtained by general relativity inspired methods in \cite{Fujii1} and discuss the situation in
higher codimensions.

In Section 3 we present the method of converting the second class constraints into the first class ones
by adding some new degrees of freedom. An error contained in \cite{myDirac} is corrected.

In Section 4 we review the thin layer quantization. In this approach a particle moves
between two equidistant infinite potential walls \cite{Jensen} or it is subject to some potential force
which in a proper limit makes it moving strictly along the surface \cite{Costa}.
In higher codimensions the method becomes quite involved and technical, but it also exhibits interesting
features \cite{Mitchell} related to a remarkable appearance of nontrivial gauge structures (for quantum
motion in submanifolds with non-flat normal bundles) first reported in \cite{Maraner1,Maraner2}.

\section{Dirac quantization}
Suppose we have a theory with $2N$ constraints $\phi_a$, $a=1,2,\ldots ,2N$. These
constraints are said to be of the second class if $\det\{\phi_a,\phi_b\}\neq 0$ even in a weak
sense \cite{Dirac} in Dirac terminology (on the constrained surface $\phi_a=0$, \ $\forall a$). In particular, it means that, unlike the first class constraints, they do not form a closed algebra with respect to the Poisson brackets. In such a situation
the standard replacement of Poisson brackets by commutators doesn't work as it would contradict
$\{\phi_a,\phi_b\}\neq 0$ inequality, and more complicated procedures are needed.
One possible way out is to introduce the Dirac 
brackets:
\begin{equation}
\label{Dirac}
\{ f,g\}_{\cal D}=\{ f,g\}-\sum_{a=1}^{2N}\sum_{b=1}^{2N}\{ f,{\phi}_a\}\Delta_{ab}\{ \phi_b,g\},
\end{equation}
where $\Delta_{ab}$ is the matrix inverse of $\{ \phi_a,\phi_b\}$. Now $\{\phi_1,\phi_2\}_D=0$ and we can introduce the commutators in a usual way. Dirac bracket is degenerate and does not define any symplectic manifold but it can be regarded as a Poisson
structure obtained by factorization of original Poisson bracket algebra over motions in unphysical direction
\cite{myDirac}, see the expression (\ref{proj}) for the momenta operators below.

As a simplest example, one can consider a free particle motion on $(n-1)$-dimensional sphere, $\sum\limits_{i=1}^{n}x^{2}_{i}=R^2$, in $n$-dimensional
Euclidean space. It can be considered as a system with two second class
($\{\phi_1,\phi_2\}=2{\overrightarrow x}^2\neq 0$) constraints \cite{KleSha}
\begin{equation}
\label{1const}
\phi_1\equiv\sum_{i=1}^{n}x_i^2-R^2=0,
\end{equation}
\begin{equation}
\label{2const}
\phi_2\equiv\sum_{i=1}^{n}x_ip_i=0
\end{equation}
where $p_i$ are canonical momenta. We impose the $\phi_1$ condition as a primary constraint directly at the Hamiltonian level. In the Lagrange formulation we also have a Lagrange multiplyer which would generate one more pair of second class constraints (vanishing of its canonical momentum as a primary constraint and some consistency condition involving the multiplyer itself). For the physical phase space it influences neither the Poisson structure nor the physical Hamiltonian \cite{KlaSha}. Thus we prefer not to increase the number of unphysical variables more than needed and work purely within the Hamiltonian mechanics. (Note that every new constraint enters the Dirac generalized Hamiltonian multiplied by its own new arbitrary function of time \cite{Dirac}.)

A natural question to be asked at this point is whether one could quantize the theory without any classical rearrangement before and then impose the constraint $\phi_1$ at the quantum level. It would be problematic because the secondary constraint $\phi_2$ (i.e. consistency condition concerning the time derivative of $\phi_1$)  shows up which has a non-vanishing Poisson bracket with $\phi_1$. The problem would be to find a natural selfadjoint restriction of the Hamiltonian to the physical space. It is possible in principle but complicated and requires quite some accuracy. We refer the reader to the Ref. \cite{Grundling} and proceed with the Dirac method.

A simple calculation according to (\ref{Dirac}) shows that \cite{KleSha}
\begin{equation}
\label{sph1}
\{ x_i,x_j\}_{\cal D}=0,
\end{equation}
\begin{equation}
\label{sph2}
\{ x_i,p_j\}_{\cal D}=\delta_{ij}-\frac{x_ix_j}{{\overrightarrow x}^2},
\end{equation}
\begin{equation}
\label{sph3}
\{ p_i,p_j\}_{\cal D}=\frac{1}{{\overrightarrow x}^2}(p_ix_j-p_jx_i).
\end{equation}
This algebra can be satisfied by very simple (usual) coordinate operators  $\hat x_i=x_i\hat I$ and
standard differential operators of momenta from which the normal differentiation is subtracted \cite{myDirac}:
\begin{equation}
\label{proj}
-i\hbar{\overrightarrow\bigtriangledown} \longrightarrow -i\hbar\left({\overrightarrow\bigtriangledown} - 
\frac{\overrightarrow x}{|\overrightarrow x|}
\left (\frac{\overrightarrow x}{|\overrightarrow x|} \cdot {\overrightarrow\bigtriangledown}\right )\right)
\equiv{\hat {\overrightarrow p}}
.\end{equation}
The problem is that $\hat p_i$ are not selfadjoint. But at the sacrifice of Leibnitz rule we can introduce  new selfadjoint momenta:
$${\hat{\tilde p}}_i=\frac{1}{2}({\hat p}_i+{\hat p}_i^{\dag})=
{\hat p}_i+i\hbar\frac{n-1}{2}\cdot\frac{x_i}{{\overrightarrow x}^2}{\hat I}.$$
The primary constraint (\ref{1const}) defines the space of physical states and the secondary one (\ref{2const}) acquires the form of identity
${\hat \phi}_2=\sum\limits_{i=1}^n({\hat x}_i{\hat{\tilde p}}_i+({\hat x}_i{\hat{\tilde p}}_i)^{\dag})\equiv 0$.
The resulting Hamiltonian \cite{myDirac}
$${\hat H}^{\cal(D)}\equiv\frac{1}{2}\sum\limits_{i=1}^n{\hat{\tilde p}}_i^2=-\frac{\hbar^2}{2}\Delta_{LB}+\frac{\hbar^2(n-1)^2}{8R^2}$$ contains a typical
quantum potential $V_q^{\cal(D)}=\frac{\hbar^2(n-1)^2}{8R^2}$.
We should stress that this result can be obtained in a purely algebraic manner without any kind of explicit operator realization \cite{KleSha}.

Note also that the same procedure may lead \cite{myDirac} to Podolsky theory if one takes our initial definition
(\ref{proj}) for
$\hat p_i$ and
Hamiltonian ${\hat H}^{\cal(P)}=\frac{1}{2}\sum\limits_{i=1}^n{\hat p}^{\dag}_ip_i$ which equals 
$-\frac{\hbar^2}{2}\Delta_{LB}$ for the physical sector functions. The quantum potential is zero:
$V_q^{\cal(P)}=0$. Thus one preserves an important property of momenta operators, the Leibnitz rule, so that they are
differentiations on the algebra of smooth functions. These operators are not selfadjoint and can't represent observables.
But in any case they do not have any clear physical meaning being projections of generators of motions along the coordinate lines of 
n-dimensional flat space, which are somewhat esoteric for an observer living on the sphere.
Natural observables on the sphere are generators of $SO(n)$ rotations, and they are selfadjoint
(proportional to $i[{\hat p}_i,{\hat p}_j]$).

\subsection{Arbitrary codimension 1 surfaces}
The free motion on a codimension 1 surface $f(x)=0$ can be obtained \cite{KlaSha} by an obvious modification
of (\ref{1const}) and (\ref{2const}):
\begin{equation}
\label{arbcon1}
\phi_1\equiv f(x)=0,
\end{equation}
\begin{equation}
\label{arbcon2}
\phi_2\equiv\sum_{i=1}^n(\partial_if)p_i=0.
\end{equation}
If $\left|{\overrightarrow \bigtriangledown}f\right|\ne0$ at the physical surface (as we assume throughout the paper), these constraints are of the second class because $\{\phi_1,\phi_2\}=\left({\overrightarrow \bigtriangledown}f\right)^2$. Again we
introduce the Dirac brackets by the prescription (\ref{Dirac}) and get
\begin{equation}
\label{arb1}
\{ x_i,x_j\}_{\cal D}=0,
\end{equation}
\begin{equation}
\label{arb2}
\{ x_i,p_j\}_{\cal D}=\delta_{ij}-\frac{(\partial_if)(\partial_jf)}{\left({\overrightarrow \bigtriangledown}f\right)^2},
\end{equation}
\begin{equation}
\label{arb3}
\{ p_i,p_j\}_{\cal D}=\frac{1}{\left({\overrightarrow \bigtriangledown}f\right)^2}\sum_{k=1}^{n}\left((\partial_jf)(\partial^2_{ik}f)-
(\partial_if)(\partial^2_{jk}f)\right)p_k.
\end{equation}
One can use the following operators \cite{myDirac} for the quantum description: ${\hat x}_i=x_i{\hat I}$,
$${\hat p}_i=-i\hbar\left (\frac{\partial}{\partial x_i}-\frac{(\partial_if)}{\left|{\overrightarrow\bigtriangledown}f\right|}
\sum\limits_{j=1}^n\frac{(\partial_jf)}{\left|{\overrightarrow\bigtriangledown}f\right|}\frac{\partial}{\partial x_j}\right )$$
as non-selfadjoint momenta and 
\begin{equation}
\label{new}
{\hat{\tilde p}}_i={\hat p}_i+
\frac{i\hbar}{2}\sum_{j=1}^n\left(\frac{\partial}{\partial x_j}\left(
\frac{(\partial_if)(\partial_jf)}{\left({\overrightarrow \bigtriangledown}f\right)^2}\right)\right)
\end{equation}
for the selfadjoint counterparts.
(Here we implement the factorization over unphysical motions again.)
The operator ordering problem is relevant only for the momenta commutators, and it is solved by
our explicit choice of the operators as follows:
\begin{equation*}
[{\hat p}_i,{\hat p}_j]=\frac{i\hbar}{\left({\overrightarrow \bigtriangledown}f\right)^2}\sum_{k=1}^{n}\left((\partial_jf)(\partial^2_{ik}f)-
(\partial_if)(\partial^2_{jk}f)\right){\hat p}_k;
\end{equation*}
\begin{equation*}
[{\hat{\tilde p}}_i,{\hat{\tilde p}}_j]=\frac{i\hbar}{2}\sum_{k=1}^n
\left (\frac{(\partial_jf)(\partial^2_{ik}f)-
(\partial_if)(\partial^2_{jk}f)}{\left({\overrightarrow \bigtriangledown}f\right)^2}{\hat{\tilde p}}_k
+{\hat{\tilde p}}_k
\frac{(\partial_jf)(\partial^2_{ik}f)-
(\partial_if)(\partial^2_{jk}f)}{\left({\overrightarrow \bigtriangledown}f\right)^2}\right).
\end{equation*}
We have the identity $\sum\limits_{i=1}^n(\partial_if){\hat p}_i\equiv 0$ \ or \
$\sum\limits_{i=1}^n\left((\partial_if){\hat {\tilde p}}_i+{\hat {\tilde p}}_i(\partial_if)\right)\equiv 0$ for the secondary constraint. And the physical sector is defined by the primary one:
$\Psi_{phys}=\psi(x)\delta(f(x))$.

For non-selfadjoint momenta the Hamiltonian reads 
$${\hat H}^{\cal(P)}=\frac{1}{2}\sum\limits_{i=1}^n{\hat p}_i^{\dag}{\hat p}_i=
-\frac{\hbar^2}{2}\left({\tilde\Delta}-\left(\frac{\partial}{\partial\overrightarrow n}\right)^2-{\rm{div}}({\overrightarrow n})\cdot\frac{\partial}{\partial\overrightarrow n}\right)$$
where $\tilde\Delta$ is the Laplace operator in the Euclidean space and ${\overrightarrow n}=\frac{\overrightarrow\bigtriangledown f}
{\left|{\overrightarrow\bigtriangledown f}\right|}$ is a unit vector normal to the surface (\ref{arbcon1}). 
In the selfadjoint case
the Hamiltonian 
${\hat H}^{\cal (D)}=\frac{1}{2}\sum\limits_{i=1}^n{\hat{\tilde p}}_i^2={\hat H}^{\cal (P)}+V^{\cal (D)}_q(x)$
contains also a quantum potential
\begin{multline}
\label{potential}
V^{\cal (D)}_q=
-\frac{\hbar^2}{8}\sum_{i=1}^n\left(\sum_{j=1}^n\frac{\partial}{\partial x_j}
\frac{(\partial_if)(\partial_jf)}{\left({\overrightarrow \bigtriangledown}f\right)^2}\right)^2+\\
+\frac{\hbar^2}{4}\sum_{i=1}^n\left(\frac{\partial}{\partial x_i}-
\sum_{k=1}^n
\frac{(\partial_if)(\partial_kf)}{\left({\overrightarrow \bigtriangledown}f\right)^2}
\frac{\partial}{\partial x_k}\right)
\left(\sum_{j=1}^n\frac{\partial}{\partial x_j}\frac{(\partial_jf)(\partial_if)}{\left({\overrightarrow \bigtriangledown}f\right)^2}\right).
\end{multline}
Some details of rather straightforward  calculations can be found in \cite{myDirac}. 
Note that in terms of the normal vectors ${\overrightarrow n}=\frac{\overrightarrow\bigtriangledown f}
{\left|{\overrightarrow\bigtriangledown f}\right|}$
one can easily write the potential (\ref{potential}) down in the following form:
\begin{equation}
 \label{normalform}
V^{\cal (D)}_q=\frac{\hbar^2}{4}\left(\frac12 \left(\sum_{i}\partial_i n_i\right)^2+
\sum_{i,k} \partial_i \left(n_k \partial_k n_i\right)+\frac12
\sum_{i,k,m}n_i n_k n_m\partial^2_{km}n_i\right)
\end{equation}
using the obvious relations $\sum\limits_i n_i\partial_k n_i=0$ and $\sum\limits_i n_i\partial^2_{km} n_i=-
\sum\limits_i \left(\partial_m n_i\right)\left(\partial_k n_i\right)$.

Unfortunately both Hamiltonians, ${\hat H}^{\cal (D)}$ and ${\hat H}^{\cal (P)}$, are ambiguous; they take
different values for those functions which represent one and the same surface.
(And the problem exists even for spheres.) Indeed, any surface can be represented by its tangent paraboloid at some point:
$f(y)=y_n-\frac{1}{2}\sum\limits_{\alpha=1}^{n-1}k_{\alpha}y_{\alpha}^2+{\cal O}(y_{\alpha}^3)$ where
$y_{\alpha}$ are Cartesian coordinates. (It is not a priori obvious that this accuracy is enough for
calculating the quantum potential but one can easily check that in this case it is, see \cite{myDirac}.)
Then (\ref{potential}) gives
$$V_q=\frac{\hbar^2}{8}\left(\left(\sum\limits_{\alpha=1}^{n-1}k_{\alpha}\right)^2+2\sum\limits_{\alpha=1}^{n-1}k_{\alpha}^2\right)+{\cal O}(y_{\alpha})$$
in the vicinity of the point $\overrightarrow y=0$.
For a sphere the principal curvatures are $k_{\alpha}=\frac{1}{R}$ and at the chosen point we have $V_q=\frac{\hbar(n^2-1)}{8R^2}$ which differs
from our previous result (and that of \cite{KleSha}). So, the Dirac recipe is ambiguous.
To fix the freedom, we propose the following (geometrically natural) choice of the function $f(x)$: up to the sign it should be equal to (some function of) the distance from the surface $f=0$. 
After that we have
$\left|\overrightarrow\bigtriangledown f\right|=1$ and $\partial_i n_k=\partial_k n_i$, 
$\sum\limits_{k}n_k\partial_kn_i=0$ where $n_k=\partial_k f$. 
Then a simple calculation \cite{myDirac} shows that
the quantum potential on the $f(x)=0$ surface is
$V^{\cal (D)}_q=\frac{\hbar^2}{8}\left(\sum\limits_{\alpha=1}^{n-1}k_{\alpha}\right)^2$.
For spheres it yields the previous result.
The kinetic part of the Hamiltonians for our choice of unit normals
$${\tilde\Delta}-\left(\frac{\partial}{\partial\overrightarrow n}\right)^2-{\rm{div}}({\overrightarrow n})\cdot\frac{\partial}{\partial \overrightarrow n}=\Delta_{LB}$$
equals to the Laplace-Beltrami operator on the physical surface \cite{myDirac}.
But in the general case (when the unit normal vector ${\overrightarrow n}=\frac{\overrightarrow\bigtriangledown f}
{\left|{\overrightarrow\bigtriangledown f}\right|}$ would not be orthogonal to the surfaces $f(x)=const\neq 0$) this result
would not be true. The vector $\overrightarrow n$ will change its direction while moving apart from the initial surface, and the second normal derivative $\left(\frac{\partial}{\partial\overrightarrow n}\right)^2$ would add some extra (first order differential) term to $\Delta_{LB}$. 

Let's consider a simple illustration. For a circle in a plane we would use
$f(x,y)=|y|-\sqrt{1-x^2}$\ (instead of $f(x,y)=x^2+y^2-1$) and approximate it near the
$(x,y)=(0,-1)$ point by a parabola $f(x)=y-\frac{x^2}{2}+1=0$. We have
$n_x=-\frac{x}{\sqrt{1+x^2}}$ and $n_y=\frac{1}{\sqrt{1+x^2}}$. The
selfadjoint momenta can be easily found as
$${\hat p}_x=-\frac{i\hbar}{1+x^2}\left(\frac{\partial}{\partial x}+x\frac{\partial}{\partial y}\right)
+\frac{i\hbar x}{\left(1+x^2\right)^2},$$
$${\hat p}_y=-\frac{i\hbar x}{1+x^2}\left(\frac{\partial}{\partial x}+x\frac{\partial}{\partial y}\right)
+\frac{i\hbar \left(x^2-1\right)}{2\left(1+x^2\right)^2}.$$
It leads to the Hamiltonian
$${\hat H}^{(\cal D)}=-\frac{\hbar^2}{2}\left(\frac{1}{1+x^2}\left(\frac{\partial}{\partial x}+x\frac{\partial}{\partial y}\right)^2-\frac{2x}{\left(1+x^2\right)^2}\left(\frac{\partial}{\partial x}+x\frac{\partial}{\partial y}\right)\right)+\frac{\hbar^2\left(3-2x^2-5x^4\right)}{8\left(1+x^2\right)^4}$$
which gives $V_q=\frac{3\hbar^2}{8}$ for the quantum potential at $x\to 0$ (instead of $\frac{\hbar^2}{8}$). And
introducing the tangent derivative $\frac{\partial}{\partial \overrightarrow t} =\frac{1}{\sqrt{1+x^2}}\left(\frac{\partial}{\partial x}+x\frac{\partial}{\partial y}\right)$ we get
$${\hat H}^{(\cal P)}=-\frac{\hbar^2}{2}\left(\Delta_{LB}-\frac{2x}{\left(1+x^2\right)^{3/2}}\cdot\frac{\partial}{\partial \overrightarrow t}\right)$$ for the kinetic energy operator with $\Delta_{LB}=\frac{\partial^2}{{\partial \overrightarrow t}^2}$.

\subsection{Relation to GR-like methods}
Some time ago this problem was tackled in Ref. \cite{Fujii1} by methods typical to general relativity. The
quantization was performed in Cartesian coordinates but a curvilinear coordinate system was also used.
One of the coordinates $q_0$ was chosen to be the value of the function $f$ and the others had to be orthogonal to 
it. Then the authors of \cite{Fujii1} defined the curvilinear momenta
${\hat p}_{\mu}$ in terms of Cartesian ones. The Cartesian commutators (obtained from Dirac brackets)
implied the commutation relations for the curvilinear operators $[q^{\nu},{\hat p}_{\mu}]
=i\hbar\left(\delta^{\nu}_{\mu}-n^{\nu}n_{\mu}\right)$ and set $p_0=0$. After doing this one can
arrive at \cite{Fujii1}
$${\hat H}^{\cal (D)}=\frac12 \sum_{i,j=1}^{n-1}g^{-1/4} {\hat p}_i g^{1/2} g^{ij} {\hat p}_j g^{-1/4} +V_q$$
where $g_{ij}$ is the metric on the physical surface and the quantum potential is equal to our result for the special choice of the equation of surface, $\frac{\hbar^2}{8}\left({\rm div}{\overrightarrow n}\right)^2$.

The authors interpreted the kinetic part of the Hamiltonian as the Laplace-Beltrami operator
on the hypersurface $f=0$. It would indeed be true if the momenta were the standard ones for the hypersurface, but they were
defined to be symmetric in the ambient space which is not the same. Up to the setting the normal differentiation to zero, they would be the standard symmetric momenta operators for the curvilinear coordinate system
in the ambient Euclidean space. It amounts to the difference between $\frac{1}{g^{1/4}}\partial_i g^{1/4}$ and $\frac{\sqrt{|\bigtriangledown f|}}{g^{1/4}}\partial_i \frac {g^{1/4}}{\sqrt{|\bigtriangledown f|}}$. If the function
$f$ depends only on the distance from the surface $\left(\partial_i |\bigtriangledown f|=0\right)$
then it makes nothing
and our results coincide. What changes if we have another function $f$? A linear differential operator
$\frac{\hbar^2}{2}\sum\limits_{i,j}\frac{\partial_i |\bigtriangledown f|}{|\bigtriangledown f|}g^{ij}\partial_j$ gets
added to $-\frac{\hbar^2}{4}\Delta_{LB}$  and two additional
terms appear in the quantum potential: $-\frac{\hbar^2}{8}\cdot\frac{\left({\overrightarrow\bigtriangledown} \left(|\bigtriangledown f|\right)\right)^2}{\left(\bigtriangledown f\right)^2}+
\frac{\hbar^2}{4}{\rm div}\frac{{\overrightarrow\bigtriangledown} \left(|\bigtriangledown f|\right)}{|\bigtriangledown f|}$. All the operations are related to the hypersurface $f=0$. In particular, 
if the gradients are taken in some Cartesian coordinates of the ambient space, the differentiations should be projected
to the hypersurface. In this coordinates the vector $\frac{{\overrightarrow\bigtriangledown} \left(|\bigtriangledown f|\right)}{|\bigtriangledown f|}$ in the tangent space of $f=0$ would have the
following components: $\frac{1}{|\bigtriangledown f|}\left(\partial_i - n_i\sum\limits_k n_k\partial_k\right)
|\bigtriangledown f|=\sum\limits_j \frac{\partial_j f}{|\bigtriangledown f|}\partial_j 
\frac{\partial_i f}{|\bigtriangledown f|}$. And the new terms in the quantum potential can
be transformed to $\frac{\hbar^2}{4}\sum\limits_{i,j}\partial_i\left(n_j\partial_j n_i\right)-
\frac{\hbar^2}{8}\sum\limits_i\left(\sum\limits_j n_j\partial_j n_i\right)^2$ in a complete accordance with (\ref{normalform}).
The differential operator converts to $\sum\limits_{i,k}n_k (\partial_k n_i)\partial_i$ in Cartesian coordinates, which is exactly what would come out of the second normal derivative
 $\left(\frac{\partial}{\partial\overrightarrow n}\right)^2$ in our approach.

\subsection{Dirac quantization in higher codimensions}
Going to higher codimensions (the surface $f^{(a)}=0$ for $a=1,2\ldots N$) complicates the things considerably because the Poisson brackets of
$\sum\limits_i p_i\partial_i f^{(a)}$ and $\sum\limits_i p_i\partial_i f^{(b)}$ (the secondary constraints) do not vanish even if 
${\overrightarrow\bigtriangledown}f^{(a)}\cdot {\overrightarrow\bigtriangledown}f^{(b)}=0$. It
affects the momenta commutators. Nevertheless, if the normal bundle of the submanifold is flat
and we can (locally) choose the set of functions $f$ satisfying the conditions $\left|{\overrightarrow\bigtriangledown}f^{(a)}\right|=1$
for all $a$ and ${\overrightarrow\bigtriangledown}f^{(a)}\cdot {\overrightarrow\bigtriangledown}f^{(b)}=0$
for $a\neq b$ then quite bulky but straightforward calculations show that
the generalization of the previous results is very simple:
$${\hat{\tilde p}}_i=-i\hbar\left (\frac{\partial}{\partial x_i}-\sum_{a=1}^{N} n_i^{(a)}
\sum\limits_{j=1}^n n_j^{(a)}\cdot\frac{\partial}{\partial x_j}\right )+
\frac{i\hbar}{2}\sum_{j=1}^n \frac{\partial}{\partial x_j}\left(\sum_{a=1}^{N}
n_i^{(a)} n_j^{(a)}\right);$$
$${\hat H}^{\cal (D)}=-\frac{\hbar^2}{2}\Delta_{LB}+\sum_{a=1}^{N} \frac{\hbar^2}{8}\left({\rm div}{\overrightarrow n^{(a)}}\right)^2.$$
It gives the extrinsic mean curvature squared for the quantum potential.

And in the general case of higher codimensions,  the
straightforward approach equations become almost untractable. But the structure of terms still
remains quite understandable. For example, in the case of codimension 2 one can check that
the momentum operators can be obtained by projecting the $\frac{\partial}{\partial x_i}$
vectors to the subspace orthogonal to both ${\overrightarrow\bigtriangledown} f^{(1)}$ and
${\overrightarrow\bigtriangledown} f^{(2)}$ according to
$${\overrightarrow V}_{\perp}={\overrightarrow V}-\frac{
{\overrightarrow a}_1\left({{\overrightarrow a}_2}\right)^2
\left({\overrightarrow a}_1{\overrightarrow V}\right)+
{\overrightarrow a}_2\left({{\overrightarrow a}_1}\right)^2
\left({\overrightarrow a}_2{\overrightarrow V}\right)-
\left({\overrightarrow a}_1{\overrightarrow a}_2\right)\left({\overrightarrow a}_2
\left({\overrightarrow a}_1{\overrightarrow V}\right)+
{\overrightarrow a}_1
\left({\overrightarrow a}_2{\overrightarrow V}\right)\right)}{{\left({{\overrightarrow a}_1}\right)^2\left({{\overrightarrow a}_2}\right)^2-
\left({\overrightarrow a}_1{\overrightarrow a}_2\right)^2}}.$$
At this point the methods of previous subsection \cite{Fujii1} become very useful. In \cite{Fujii2} it
has been shown that these methods lead to formally the same result as for the codimension 1 case
(Laplace-Beltrami operator and extrinsic mean curvature squared). As it was above, the interpretation of
the kinetic energy operator is correct if the normal components of the metric do not influence
the momenta operators which means that ${\rm det}\left({\overrightarrow\bigtriangledown}f^{(a)}\cdot {\overrightarrow\bigtriangledown}f^{(b)}\right)$ does not depend on tangential coordinates.

\section{Abelian conversion method}
Sometimes in the quantum field theory first class constraints may fail to 
form a closed algebra at the quantum level. Formally, they can be said to become second class upon
quantization, but it is very bad for the quantum field theory with local symmetries as it corresponds to a gauge symmetry breaking. For example, this kind of anomaly occurs
for the Gauss law in the model of Weyl fermions interacting with
a Yang-Mills field.
 In  \cite{Faddeev} it was proposed to introduce  some new degree  of freedom in the model in
order to have new constraints with the Abelian algebra and get rid of the anomaly. It
resembles the appearance of the conformal factor as a new degree of freedom at
the quantum level for noncritical strings \cite{Polyakov}. (Very similar situation occurs
in the theory of relativistic branes \cite{Tucker}, but it is not so clear what to make out of it
in this context and also it is not known if the critical dimension exists, see
\cite{Scholl,Bars,hayashi} for diverse views on the subject.)

This idea can also be used in quantum mechanics with second class constraints \cite{KleSha}.
The (``Abelian conversion'') method consists of introducing new canonical pair of variables
$Q,\ K$ and first class constraints $\sigma_1,\ \sigma_2$: $\{\sigma_1,\sigma_2\}=0$ (identically,
so that the constraint algebra is Abelian) assuming
$\sigma_1=\phi_1,\ \sigma_2=\phi_2$ if $Q=0$ and $K=0$. In our case it would be
$\sigma_1=f(x)+K$ and $\sigma_2=\overrightarrow n\cdot\overrightarrow p+\left|{\overrightarrow\bigtriangledown}f\right|\cdot Q$. (Note an error at this point in \cite{myDirac} where the $\left|{\overrightarrow\bigtriangledown}f\right|$ factor
has disappeared from $\sigma_2$.) The next step
is to find a new Hamiltonian such that $H_S=H$ if $Q=0$ and $K=0$ and 
$\{H_S,\sigma_1\}=\{H_S,\sigma_2\}=0$. The physical sector is obtained by setting
$\sigma_1=\sigma_2=0$.
For free motion on spheres this method gave zero quantum potential, see \cite{KleSha,KlaSha}. 
Actually, the authors of \cite{KleSha,KlaSha} had the result of the form
$H_S=H_S\left(\sigma_1,\sigma_2,\sum\limits_{i<k}(x_ip_k-x_kp_i)^2\right)$ which due to relation
 $$\sum\limits_{i<k}\left(x_ip_k-x_kp_i\right)^2=\left(\sum\limits_ix_i^2\right)\left(\sum\limits_ip_i^2
-\left(\sum\limits_in_ip_i\right)^2\right)$$ could be transformed to $H_{phys}=\frac{{\overrightarrow p}^2_{phys}}{2}$ because
$\sigma_2^2=(\sum\limits_in_ip_i)^2$ if $Q=0$.

It was concluded that the Abelian conversion method is preferable because it 
involves no extrinsic geometry in the results. For this and other reasons it was used
in the beautiful projection operator approach to path integral quantization of constrained systems
\cite{Klauder1} aiming at quantizing the gravity \cite{Klauder2}. But is it possible to generalize the above result
to other surfaces? In principle, methods which introduce new variables are very strong, see for example
\cite{Batalin} and references therein. Moreover, one can proceed with a general philosophy of
Abelian conversion \cite{KlaSha} without any need of introducing such weird concepts like ghost operators \cite{Batalin}. The question is whether  it is possible to do this using the very simple geometric form
of the new constraints introduced above and obtaining some relatively simple and physically sensible results.
Let us search for $H_S=H_S(\sigma_1,\sigma_2,x,p)$ in the previous form
\begin{equation}
\label{sphlike}
 H_S=H_S\left(\sigma_1,\sigma_2,g(x)\left(\sum\limits_ip_i^2
-\left(\sum\limits_in_ip_i\right)^2\right)\right)
\end{equation}
 designed for getting the pure Laplace-Beltrami solution. The following relations show up:
\begin{eqnarray}
\label{relat1}
\{H_S,\sigma_1\}=-\sum_in_i\frac{\partial H_S}{\partial p_i}=0,\\
\label{relat2}
\{H_S,\sigma_2\}=\sum_in_i\frac{\partial H_S}{\partial x_i}-\sum_{i,k}p_k(\partial_i n_k)\frac{\partial H_S}{\partial p_i}-\sum_i\left(\partial_i\left|{\overrightarrow\bigtriangledown}f\right|\right)Q\frac{\partial H_S}{\partial p_i}=0.
\end{eqnarray}
From (\ref{relat2}) we have
\begin{multline*}
\sum_{i,k}p_ip_k\left(\sum_j n_j\partial_jg(x)(\delta_{ik}-n_in_k)-2g(x)(\partial_in_k)\right)-\\ -
2\sum_i\left(\partial_i\left|{\overrightarrow\bigtriangledown}f\right|\right)Q\left(p_i-n_i\sum_j n_j p_j\right)=0.
\end{multline*}
If we admit the condition $\left|{\overrightarrow\bigtriangledown}f\right|=1$ (compare with the Dirac method above) then
the last term disappears and
it has a non-zero solution for spheres because
$\partial_in_k\sim\delta_{ik}-n_in_k$. But this is not
true for arbitrary surfaces. Hence the result of \cite{KleSha} can't be
generalized directly.
Nevertheless, using a simple ansatz
$$H_S=H_S\left(\sigma_1,\sigma_2,\sum_{i,k}C_{ik}(x)p_ip_k+\sum_iD_i(x) p_i+E(x)\right)$$
with $C_{ik}=C_{ki}$ one can show \cite{myDirac} that in general it 
is possible to get a quadratic in momenta Hamiltonian (not equal to the Laplace-Beltrami operator) by this method if we admit the above definition of the function $f(x)$.

If another function is used then (\ref{relat2}) can't be valid identically because the last term contains $Q$.
But we can afford having it only in a weak sense. Then from $\sigma_2$ we determine 
$Q=-\frac{\overrightarrow n\cdot\overrightarrow p}{\left|{\overrightarrow\bigtriangledown}f\right|}$, and 
(\ref{relat2}) with the ansatz (\ref{sphlike}) converts into
$$\sum_{i,k}p_ip_k\left(\sum_j n_j\partial_jg(x)\left(\delta_{ik}-n_in_k\right)+
2g(x)\left(n_i \sum_j n_j \partial_j n_k-\partial_i n_k\right)\right)=0$$
which is probably not as hopeless as it was erroneously stated in \cite{myDirac} due to the
aforementioned mistake but still can give no guarantee for the existence of a non-trivial solution.
Another problem of the method is that in this setting there is no clear reason for insisting on
the Abelian algebra of the first class constraints. And if we go to higher codimensions we would really need 
to modify the method somehow, at least by finding a more clever choice of the new constraints because,
as we already mentioned, the Poisson brackets of
$\sum\limits_i p_i n_i^{(a)}$ and $\sum\limits_i p_i n_i^{(b)}$ do not vanish even if 
${\overrightarrow n}^{(a)}\cdot {\overrightarrow n}^{(b)}=0$.

\section{Thin layer quantization method}
As it was discussed in the Section 2, imposing the second class constraints directly at the
quantum level is problematic. Nevertheless, we can use a more delicate procedure. We can approximate the constrained system by a motion in a thin layer around it.
In quantum mechanics this approach appeared in \cite{Jensen,Costa}, for a deeper
discussion see \cite{Mitchell,Jaffe} and \cite{Herbst} at the mathematical level. It can also
be used for classical systems \cite{Herbst} and gives rise to extra potential if we take
the initial conditions involving normal motions in the thin tube, but coincides with the intrinsic description
if the initial velocities are tangential. However, in quantum mechanics this approach always gives a geometric potential
because in the quantum realm it is impossible to eliminate the normal motion completely.

We consider $(n-1)$-dimensional smooth surface in ${\mathbb R}^n$ and two infinite potential walls
 at the distance $\delta\to 0$ from the surface. Free quantum particle moves in the thin layer of width
 $2\delta$ between these potential walls. We introduce a curvilinear coordinate system in which
 $|x_n|$ equals the distance from the surface to the given point (thus playing the same role
as the function $f$ in the refined approach to Dirac quantization), and the coordinate lines
 of $x_1,\ldots,x_{n-1}$ are orthogonal to that of $x_n$. We have the boundary condition
 $\left.\Psi\right|_{x_n=\delta}=\left.\Psi\right|_{x_n=-\delta}=0$ and Hamiltonian 
 ${\tilde H}=-\frac{\hbar^2}{2}{\tilde\Delta}$ with $\tilde\Delta$ being the Laplace operator,
 $${\tilde\Delta}=\sum\limits_{i=1}^n\sum\limits_{k=1}^n{\tilde g}^{-1/2}\partial_i{\tilde g}^{1/2}{\tilde g}^{ik}\partial_k
={\partial_n}^2+\left({\tilde g}^{-1/2}\partial_n{\tilde g}^{1/2}\right)\partial_n+\Delta_{LB},$$
$${\tilde g}_{ik}=\left (
\begin{matrix}
g_{ab}&0 \\ 
0&1
\end{matrix}
\right)$$
where $\Delta_{LB}$ is the Laplace-Beltrami operator on the surface $x_n=const$.

The simplest way \cite{mythin} to obtain the thin layer limit is to consider
the tangent paraboloid of the surface
$y_n=\frac{1}{2}\sum\limits_{a=1}^{n-1}k_ay_a^2+{\cal O}(y_a^3)$, where
$k_a$ are the principal curvatures. The unit normal is
$n_a=\frac{k_ay_a}{\sqrt{1+\sum\limits_{a=1}^{n-1}k_a^2y_a^2}}+
{\cal O}(y_a^2)=k_ay_a+{\cal O}(y_a^2)$, $n_n=-1+{\cal O}(y_a^2)$ and
\begin{equation}
\label{div}
{\rm{div}}{\overrightarrow n}=\sum_{a=1}^{n-1}k_a+{\cal O}(y_a).
\end{equation}
A nearby surface $x_n=\epsilon$ can be obtained by taking
${\overrightarrow y}\longrightarrow{\overrightarrow y}^{\prime}=
{\overrightarrow y}+\epsilon\overrightarrow n$ and
$dy_a^{\prime}=dy_a\left(1+\epsilon k_a+{\cal O}(y_a)\right)$. 
It yields $\frac{dS^{\prime}}{dS}=\frac{\prod\limits_{a=1}^{n-1}\left(1+{\cal O}(y_a^{\prime 2})\right)dy_a^{\prime}}
{\prod\limits_{a=1}^{n-1}\left(1+{\cal O}(y_a^{2})\right)dy_a}=\prod\limits_{a=1}^{n-1}
(1+\epsilon k_a)+{\cal O}(y_a)$ near the point $\overrightarrow y=0$. At the line
$y_a=0 \quad\forall a=1,\ldots,n-1$ one has
\begin{equation}
\label{areas}
\frac{dS^{\prime}}{dS}=1+\epsilon\sum_{a=1}^{n-1}k_a+\frac{1}{2}\epsilon^2\left(
\left(\sum_{a=1}^{n-1}k_a\right)^2-\sum_{a=1}^{n-1}k_a^2\right)+{\cal O}(\epsilon^3).
\end{equation}
Clearly, the relation (\ref{areas}) is valid for every point of the surface with its own principal curvatures.
Following  \cite{Jensen,Costa} we introduce a new wave function $$\chi (x)=\Psi (x)\sqrt{\frac{dS^{\prime}}{dS}}.$$
Physically it amounts to $$\int\limits_{|x_n|\leq\delta}dV|\Psi (x)|^2=\int\limits_{-\delta}^{\delta}dx_n
\int dS|\chi (x)|^2,$$ 
so that the function $\int dx_n|\chi (x)|^2$ defines the probability density of finding the particle at a given point on the surface. For the lowest
energy solution the normal motion gives only the factor of $\cos\frac{\pi x_n}{2\delta}$,
and we
easily get \cite{mythin}
\begin{equation*}
{\tilde\Delta}\Psi (x)=
{\Delta}_{LB}\chi (x)+\partial_n^2\chi(x)+
\left(\frac{1}{2}\sum_{a=1}^{n-1}k_a^2-\frac{1}{4}\left(\sum_{a=1}^{n-1}k_a\right)^2\right)\chi(x)
+{\cal O}(x_n).
\end{equation*}
At this  energy level
$\chi(x_1,\ldots,x_n)=f(x_1,\ldots,x_{n-1})\cos\frac{\pi x_n}{2\delta}$.
After taking $\delta\to 0$ limit and subtracting an infinite (proportional to
$1/{\delta^2}$) energy we obtain the Hamiltonian
\begin{equation}
\label{Hamilton}
{\hat H}=-\frac{\hbar^2}{2}\Delta_{LB}+\frac{\hbar^2}{8}
\left(\left(\sum_{a=1}^{n-1}k_a\right)^2-2\sum_{a=1}^{n-1}k_a^2\right)
\end{equation}
which contains the quantum potential
$$V_q=\frac{\hbar^2}{8} 
\left(\left(\sum_{a=1}^{n-1}k_a\right)^2-2\sum_{a=1}^{n-1}k_a^2\right).$$
For 2-dimensional surfaces in ${\mathbb R}^3$ the result of da Costa \cite{Costa},
$V_q=-\frac{\hbar^2}{8}(k_1-k_2)^2$, is reproduced; for spheres $k_a=\frac{1}{R}$ and the potential
is $V_q=\frac{\hbar^2(n-1)(n-3)}{8R^2}$. If we would use a layer of varying width, some additional
effective forces will appear \cite{Mitchell}.

\subsection{Some remarks and variations}
We could use an appropriate
confining potential instead of infinite walls. It would lead to 
the lowest energy level function of the potential
$V_{conf}(\frac{x_n}{\delta})$ instead of $\cos\frac{\pi x_n}{2\delta}$ and to
another infinite energy. Note that we can also embed one curved space into another curved space \cite{Mitchell}.
Moreover, this approach can be used for the quantum graphs theory \cite{Tenuta}. And it is currently used
to describe a motion of electrons in nanostructures \cite{EnEti,Mott} and for the physics of molecules
\cite{Mitchell,Maraner}. Even before the general consideration of the problem appeared in \cite{Jensen}
and \cite{Costa}, some elements of the thin layer approach were successfully used in the theory of chemical reactions
\cite{Marcus}. Recently, a considerable progress in the method has been achieved \cite{Teufel} allowing one to treat a very general type of quantum constrained motion, even with relatively large kinetic energies, with a full mathematical rigour.

We should mention that there is one more method of quantization proposed
by Prokhorov \cite{Nuramatov}. The motion of a particle is considered as a system with two
second class constraints but only one condition is imposed on the physical sector:
${\hat P}_n\Psi_{phys} (x)=0$ with ${\hat P}_n=-i\hbar
\frac{1}{{\tilde g}^{1/4}}\frac{\partial}{\partial x_n}{\tilde g}^{1/4}$. It means
that
\begin{equation}
\label{Prokhorov}
\partial_n\left(\sqrt{\frac{dS^{\prime}}{dS}}\Psi_{phys}(x)\right)=0.
\end{equation}
Having solved some task by this method, one should put $x_n=0$ in the results
{\it after} all the differentiations over $x_n$ are performed. 
Due to (\ref{Prokhorov}) the probability to find a particle at the distance $|x_n|$ from
the surface does not depend on the value of $x_n$, and we choose one value
we need. (For Prokhorov's view see \cite{Nuramatov}.)
This method gives the same results \cite{mythin} as the thin layer approach due to a
very simple physical reason. The lowest energy level wave functions (in the model with
two infinite potential walls) have nodes at $x_n=\pm\delta$ and the bunch at
$x_n=0$: $\partial_n\chi=0$ or, equivalently, ${\hat P}_n\Psi=0$.

One also could be tempted to use the Hamiltonian in curvilinear coordinates 
$H=\frac12\left(\sum\limits_{i=1}^{n-1}\frac{p_i^2}{h_i(x)}+p_n^2\right)$ with a simple recipe
$p_i\longrightarrow{\hat p_i}=-i\hbar{\tilde g}^{-1/4}\partial_i{\tilde g}^{1/4}$
followed by the thin layer method. Then the result \cite{Encinosa} for $\psi(x_1,\ldots,x_n)=f(x_1,\ldots,x_{n-1})\cdot\cos\frac{\pi x_n}{2\delta}$
\begin{equation*}
{\hat H}\frac{f\cdot\cos\frac{\pi x_n}{2\delta}}{\sqrt{\frac{dS^{\prime}}{dS}}}=
\frac12\sum\limits_{i=1}^{n-1}\frac{\hat p_i^2}{h_i(x)}\frac{f\cdot\cos\frac{\pi x_n}{2\delta}}{\sqrt{\frac{dS^{\prime}}{dS}}}
-\frac12\frac{\hbar^2f}{\sqrt{\frac{dS^{\prime}}{dS}}}\ \partial_n^2\cos\frac{\pi x_n}{2\delta}.
\end{equation*}
can be considered as zero quantum potential. 
However, one should remember that quantization in curvilinear coordinates is dangerous because its results usually
depend on the choice of coordinate system. And the curvilinear
 momenta operators are only symmetric but not in general selfadjoint.
And what is more important, the operator ordering problem in
$\frac{\hat p_i^2}{h_i(x)}$ terms is not solved. It is not difficult to deduce
the correct ordering for the zero potential theory, but this particular ordering is not so natural {\it a priori} and 
can involve quite bulky expressions \cite{mythin}. (See also \cite{Liu}.)

\subsection{Higher codimensions and gauge structures}
In general we can represent a smooth $m$-dimensional surface in ${\mathbb R}^n$ by its tangent
paraboloid at a chosen point:
\begin{equation}
\label{codim}
y_{\alpha}=\frac{1}{2}\sum_{a=1}^m\sum_{b=1}^mk^{(\alpha)}_{ab}y_ay_b+{\cal O}(y_a^3),
\end{equation}
$\alpha=m+1,\ldots,n$ with some ``curvature coefficients'' $k^{(\alpha)}_{ab}=k^{(\alpha)}_{ba}$. But it turns out that this approach is convenient only for quantization on curves (and for some other relatively trivial cases
like flat 2-torus in ${\mathbb R}^4$).

\subsubsection{Quantization on curves}
For $1$-dimensional manifolds (curves) a suitable rotation in the space of $y_{\alpha}$ casts (\ref{codim}) to the form
$y_2=\frac{1}{2}ky_1^2+{\cal O}(y_1^3)$; $y_3,\ldots,y_n={\cal O}(y_1^3)$. The unit normal vectors
are $n_1^{(2)}=ky_1+{\cal O}(y_1^2)$, $n_2^{(2)}=-1
+{\cal O}(y_1^2)$, $n_3^{(2)}=\ldots=n_n^{(2)}={\cal O}(y_1^2)$; $n_i^{(\alpha)}=-\delta_{i\alpha}+
{\cal O}(y_1^2)$ for $\alpha\geq 3$. We have ${\overrightarrow n}^{(\alpha)}{\overrightarrow n}^{(\beta)}=\delta_{\alpha\beta}+{\cal O}(y_1^2)$,
and after the transformation ${\overrightarrow y}\to{\overrightarrow y}^{\prime}={\overrightarrow y}+
\sum\limits_{\alpha=2}^n\epsilon_{\alpha}{\overrightarrow n}^{(\alpha)}$ one gets
$dy_1^{\prime}=(1+\epsilon_2k+{\cal O}(y_1))dy_1$ and $dy_{\alpha}^{\prime}=(1+{\cal O}(y_1))dy_{\alpha}$ for $\alpha\geq 3$.
We 
introduce a new curvilinear coordinate system near the curve in which $x_1$ is just the length along the curve
and the hypersurfaces of constant $x_1$ are the cross sections of its tubular neighbourhood. And in a given 
cross section any point
 $\overrightarrow r$ has other $n-1$ coordinates defined by
$x_{\alpha}={\overrightarrow n}^{(\alpha)}\cdot{\overrightarrow r}$. In this coordinate system
$${\tilde g}_{ik}=\left (
\begin{matrix}
(1+x_2k)^2&0 \\ 
0&I
\end{matrix}
\right)$$ and
${\tilde\Delta}=\Delta_{c}+\Delta_{n}+\left(\frac{1}{1+x_2k}\partial_2(1+x_2k)\right)\partial_2=
\Delta_{c}+\Delta_{n}+\frac{k}{1+x_2k}\partial_2$ where $\Delta_{c}$ is Laplace-Beltrami operator
on a curve $x_{\alpha}=const$ and $\Delta_{n}=\sum\limits_{\alpha=2}^n\partial_{\alpha}^2$ is Laplace
operator in a hyperplane $x_1=const$.
Then for a wave function $\chi(x)=\sqrt{1+x_2k}\ \Psi(x)$ 
in a thin layer $\sum\limits_{\alpha=2}^nx_{\alpha}^2\leqslant\delta^2$
we obtain \cite{mythin}
\begin{equation*}
{\tilde\Delta}\Psi(x)={\tilde\Delta}\frac{\chi(x)}{\sqrt{1+x_2k}}=
{\Delta}_c\chi(x)+{\Delta}_n\chi(x)+\frac{k^2}{4}\chi(x)+{\cal O}(x_{\alpha}).
\end{equation*}
After subtracting an infinite energy due to ${\Delta}_n\chi(x)$ it yields the quantum potential
$V_q=-\frac{\hbar^2}{8}k^2$ as in \cite{Costa}. At this point we can also see that the higher codimensional thin layer problem is not reducible to a step-by-step decreasing of the physical space dimension. Indeed, for a straight line in ${\mathbb R}^3$ we obviously have zero quantum potential. But if we first consider a cylinder of radius $R$ and then restrict it further to the line, the quantum potential would be $-\frac{\hbar^2}{8R^2}$.

There is one subtlety in the above discussion. If the curve has a torsion, our coordinate system rotates around it.
If one attempts at describing this motion in a non-rotating coordinate system he gets new terms in the Hamiltonian
which correspond to rotations around the curve \cite{Takagi}. Locally these descriptions are equivalent, but globally
for a closed curve the rotating coordinate system may not exist and one can get global phases out of it \cite{Takagi}. (Global
effects may also be relevant for nontrivial normal bundles, for example for a motion on the M{\"o}bius strip,
\cite{Herbst}.) Note that in our consideration we need either to use a thin tube with spherical cross sections or to rotate the cross section around the curve together with its Frenet frame. For more general discussion of twisting the confining potential see \cite{Mitchell}.

\subsubsection{Higher dimensions and geometry of normal bundles}

In general we can try to use a similar construction; but such coordinate systems
which eliminate all the rotations from the Hamiltonian do not exist
for submanifolds with non-flat normal bundles. It was first noticed by da Costa \cite{Costa2}
who proved that it is not always possible to find a smooth family of normal vectors
with the properties required for separation of normal and tangential motions. Geometrically we need to understand the structure of a tubular neighbourhood of the physical submanifold which is naturally related to a small portion of the normal bundle corresponding to its embedding into the ambient space. Then one has to define the standard mathematical notion of the normal connection in the normal bundle \cite{Mitchell,PDE}. We will not discuss it here in any detail, but basically it amounts to taking the normal projections of the ambient space covariant derivatives. From the ambient (Euclidean) space viewpoint parallel transports according to this connection involve rotations including those which go around the physical subspace. For example, in the tubular neighbourhood of a curve the normal connection rotates the normal vectors not only with the normal hyperplanes but also around the curve together with its Frenet frame. Of course, on a curve we can exclude this effect by a suitable choice of rotating coordinates \cite{mythin} as in the previous subsubsection. But for higher dimensional (and codimensional) manifolds the normal bundle can be non-flat and such exclusion would be impossible.
The simplest examples are the configuration space of the double pendulum \cite{Costa2}
and the helical surface from \cite{Jaffe}.

Analytically it means that the Laplace operator necessarily contains terms with mixed normal and tangential derivatives, $\partial^2_{a\alpha}$. They vanish at the physical surface linearly with the distance from it, bit it is obviously not enough for the thin layer approach (unlike the Prokhorov method). Da Costa concluded \cite{Costa2} that the thin layer quantization would not work well in this situation. This statement was
repeated in \cite{mythin}.
Strictly speaking, this attitude is not right because all the dangerous terms sum up to the angular momenta operators
corresponding to rotations in normal sections of the thin tube \cite{Mitchell,Jaffe,Maraner1,Maraner2}.  In the case of the lowest energy
solutions for a thin tube with a spherical cross section it will not influence the resulting theory at all. 
But if the chosen normal energy level is degenerate then a gauge structure will show up \cite{Mitchell,Jaffe,Fujii3,Fujii4,Maraner1,Maraner2}.
The simplest example (although not very natural from the thin layer quantization viewpoint) is a higher
energy level of normal motion in the thin tube \cite{Jaffe}. In any case, the quantum potential can be
calculated explicitly \cite{Mitchell,Jaffe} but these complicated expressions do not give much to our
intuition and we omit them here. (One can attempt at making the calculations by means of the tangent paraboloid technique \cite{mythin} which was so useful above, but in the general case more powerful geometric methods \cite{Mitchell,Jaffe} are more safe and easier to implement.) Note that the Prokhorov method \cite{Nuramatov} can give the quantum
potential too \cite{mythin} but it is completely insensitive to the gauge structures.
And let us finally mention that gauge structures (different from above) appear also in the algebraic approaches to
quantization on a coset space \cite{tsu}; we do not discuss it here.

\section{Conclusions}
We presented the main approaches to quantization of systems with second class constraints which do not involve
path integrals: redefinition of the Poisson structure (Dirac brackets), conversion of the constraints
into the first class ones by introduction of new degrees of freedom and the thin tube approximation
identical to a way of realizing the holonomic constraints in classical mechanics. These methods give
different results but often they involve very similar geometric constructions and conditions. Many of the
aspects of theory deserve a better understanding. Geometric properties of the Abelian conversion
are still unclear. And it's worth trying to describe the main features of higher codimensional Dirac quantization from
an explicit operatorial perspective. This investigation can be very important for our understanding
of constrained quantum mechanics in general.

A special remark is in order concerning the reference list. We do not attempt at making it absolutely complete. But, together with the references in the references, it should suffice to give a more-or-less full picture of research in the subject, at least as much as we were able to find it out in the literature. We think it is quite important because many similar results are scattered in different works being completely disconnected from each other. Jensen and Koppe \cite{Jensen} didn't know about the article of Marcus \cite{Marcus}, and neither of these works has influenced the results of da Costa \cite{Costa}. The second article by da Costa \cite{Costa2} is rarely cited (as well as the pioneering one due to Marcus \cite{Marcus}). The authors of \cite{KleSha} and \cite{KlaSha} were unaware of earlier works \cite{Fujii1} and \cite{Fujii2}. And in our articles \cite{myDirac,mythin} we didn't refer to very important works \cite{Fujii1,Mitchell,Fujii2,Maraner,Marcus,Takagi} due to the lack of knowledge.

{\bf Acknowledgements.} The Author is greatful to the Cluster of Excellence EXC 153  ``Origin and Structure of the Universe'' for partial support; to Prof. Lev Prokhorov for pointing at the problems of quantization
with second class constraints when the Author was yet a student; to Prof. Kanji Fujii for very useful
e-mail correspondences concerning his articles on quantization problems; to the Organizers of the
40th Symposium on Mathematical Physics for the opportunity to participate in this
wonderful conference and to give a talk;
to Prof. Julio Guerrero for the interesting discussion at the Symposium.

\end{document}